%% file: eprint.tex
\documentclass[12pt]{article}
\usepackage{graphicx}

\def\institute{Department of Physics and Astronomy\\
Michigan State University}

\def\Title#1{\begin{center} {\Large #1 } \end{center}}
\def\Author#1{\begin{center}{ \sc #1} \end{center}}
\def\Address#1{\begin{center}{ \it #1} \end{center}}

\newenvironment{Abstract}{\begin{quotation}  }{\end{quotation}}
\newenvironment{Presented}{\begin{quotation} \begin{center} 
             PRESENTED AT\end{center}\bigskip 
      \begin{center}\begin{large}}{\end{large}\end{center} \end{quotation}}
\def\Acknowledgements{\bigskip  \bigskip \begin{center} \begin{large}
             \bf ACKNOWLEDGEMENTS \end{large}\end{center}}

\input econfmacros.tex

\begin{document}
\begin{titlepage}

\vfill
\Title{Using Machine Learning to Improve PDF Uncertainties}
\vfill
\Author{ Jason P.~Gombas, Reinhard Schwienhorst, Binbin Dong, and Jarrett Fein }
\Address{\institute}
\vfill
\begin{Abstract}
Parton Distribution Functions (PDFs) contribute significantly to the uncertainty on the determination of the top-quark pole mass and other precision measurements at the Large Hadron Collider (LHC). It is crucial to understand these uncertainties and reduce them to obtain the next generation of precision measurements at the LHC. The region of high momentum fraction offers an opportunity to make improvements to the PDFs. This study uses machine learning techniques in $t\bar{t}$ production to target this region of the PDF set and has potential to significantly reduce its uncertainty.
\end{Abstract}
\vfill
\begin{Presented}
$16^\mathrm{th}$ International Workshop on Top Quark Physics\\
(Top2023), 24--29 September, 2023
\end{Presented}
\vfill
\end{titlepage}
\def\thefootnote{\fnsymbol{footnote}}
\setcounter{footnote}{0}

\section{Introduction}

With the upcoming high luminosity large hadron collider (HL-LHC) \cite{ApollinariG.:2017ojx}, the next generation of precision measurements will be made. These measurements will be extremely precise, and will require lower theoretical uncertainties. Parton distribution functions (PDFs) are becoming the more dominant theoretical uncertainty in measurements like top quark pair production. However, many other measurements will also require reduced PDF uncertainties \cite{Schwienhorst:2022yqu}. 

Colliders that will offer useful data that can significantly reduce PDF uncertainty, like the Electron Ion Collider \cite{khalek2022snowmass}, are far in the future. In the meantime, PDF uncertainty can be reduced using HL-LHC data. 

Machine learning techniques can be used to pre-process the data to distill useful information to reduce specific regions of the PDFs with high uncertainty. Previous techniques involved using one, two or three dimensions of the high dimensional phase space of collider data to add to the global PDF fit \cite{Yan_2023}. Variables like rapidity and $p_Z$ of the top are typical choices. These variables do not include all the information. 
This proceeding will show current progress and some of preliminary results. 

\section{Simulation and Pseudo-Data}
A sample of $t\bar{t}$ plus one jet events with a center of mass energy of 14 TeV was generated using Madgraph at next-to-leading order (NLO) \cite{Alwall:2014hca}. A total of 7.5 million events were generated. The PDF set that was selected to study in detail was the CT18NLO PDF set \cite{Hou:2019efy}. The study looks at the truth level of $t\bar{t}j$ events without decaying the top quarks. The aim of this study is to constrain the high x region of the gluon PDF. $t\bar{t}j$ has been shown to be a process that has good potential to reduce this region of the PDFs \cite{gombas2022dependence}. 

\section{A Machine Learning Technique}
\begin{figure}
    \centering
    \includegraphics[width=0.7\textwidth]{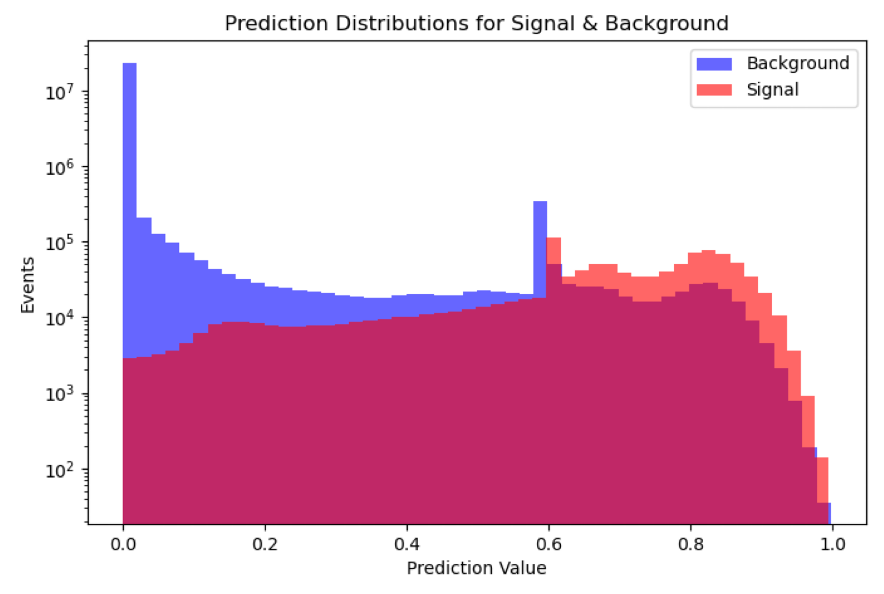}
    \caption{The MLP output scores for the trained MLP. Events that are closer to 1 are events that the MLP predicts to have an initial gluon parton whose initial momentum is greater than 2 TeV. Events that are closer to 0 are any other event. The inputs to this MLP are the kinematic 4-vectors of the final state $t\bar{t}j$. There are 3 fully connected hidden layers. The peak at about 0.6 is currently not well understood. }
    \label{fig:MLP_output}
\end{figure}

To test the idea of using machine learning to improve PDF fits, a MLP was developed to separate events with an initial gluon parton that had greater than 2 TeV longitudinal momentum. These $t\bar{t}j$ events were considered signal. Events with less than 2 TeV longitudinal momentum were considered background. The inputs to the MLP were the kinematic 4-vectors of the final state particles ($t\bar{t}j$). Decent separation was achieved which, not surprisingly, indicates that there is information about the initial colliding partons (flavor and initial momentum) in just the kinematics of the final state particles. If the MLP output score was higher than 0.7, it was considered signal and passed the MLP "filter". The MLP output scores can be seen in Fig.~\ref{fig:MLP_output}.

\section{PDF Update}
\begin{figure}[htp]
    \centering
    \includegraphics[width=0.3\textwidth]{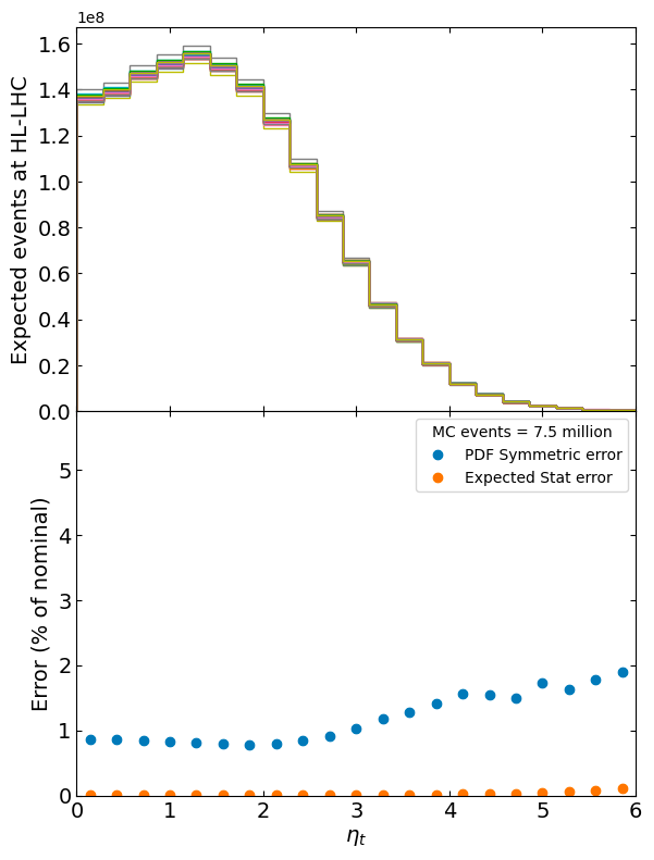}\hfill
    \includegraphics[width=0.3\textwidth]{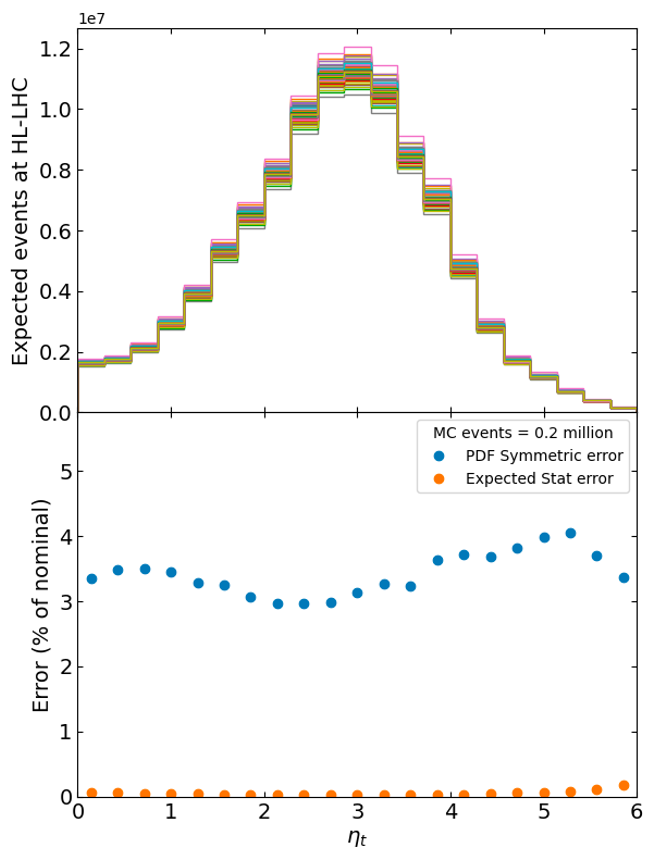}\hfill
    \includegraphics[width=0.3\textwidth]{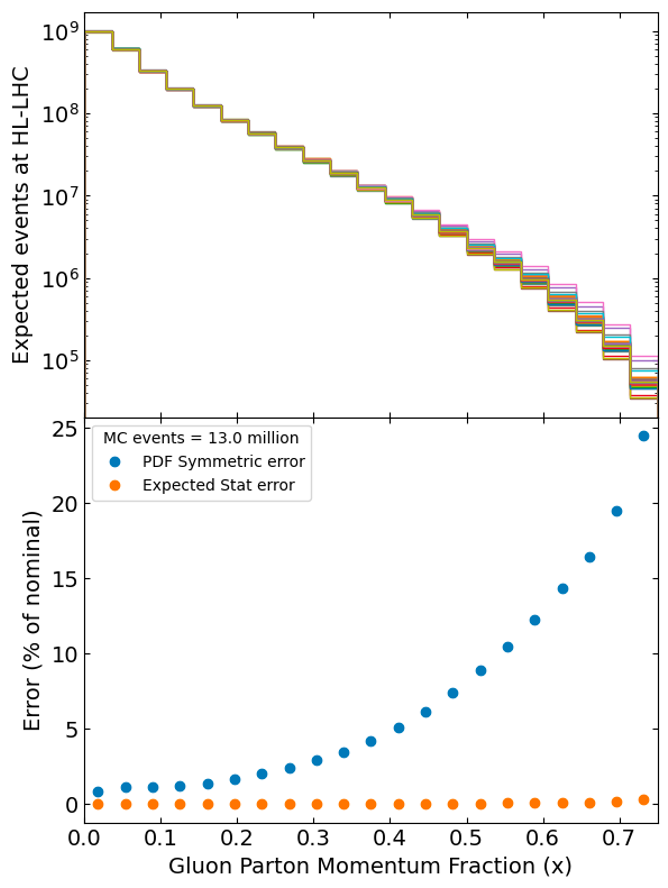}
    \caption{Differential, pseudo-data distributions that were input into ePump to update the PDF uncertainty.}
    \label{fig:epump_input}
\end{figure}

Two different differential distributions were created to compare how different methods can reduce PDF uncertainty. One histogram was filled if it passed the MLP filter and another which included every event. The rapidity of the top quark was chosen to be the kinematic variable. These differential distributions were then fed into ePump \cite{Schmidt:2018hvu} to see how much each could constrain the PDF uncertainty bands. They can be seen in Fig.~\ref{fig:epump_input}.

A systematic uncertainty of 1\% for the pseudo-data was chosen, and the statistical error was set to zero. This is because with HL-LHC data, the statistical uncertainty is expected to be negligible. This can be confirmed in Fig.~\ref{fig:epump_input} which shows that the PDF uncertainty is larger than the expected statistical uncertainty.

As a best-case-scenario, the gluon PDF is directly fed into ePump with 1\% systematic uncertainties. This provides a useful upper limit to how much the gluon PDF uncertainty can be reduced with HL-LHC data. 

\section{Results}
\begin{figure}[htp]
    \centering
    \includegraphics[width=0.3\textwidth]{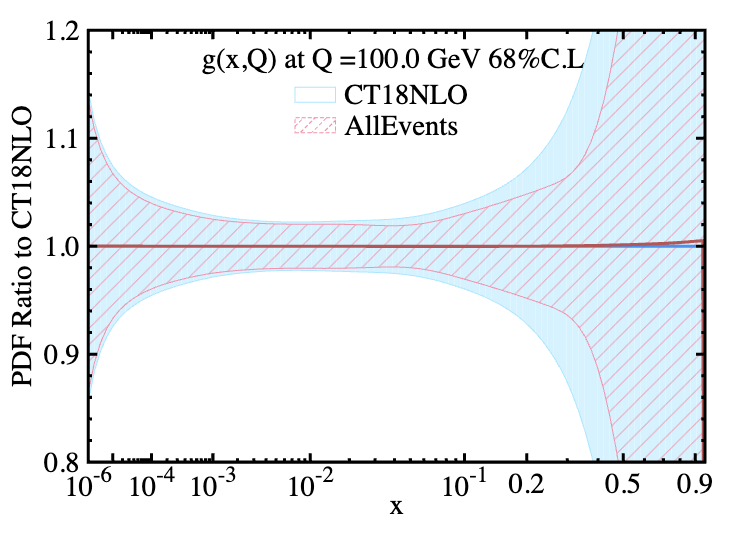}\hfill
    \includegraphics[width=0.3\textwidth]{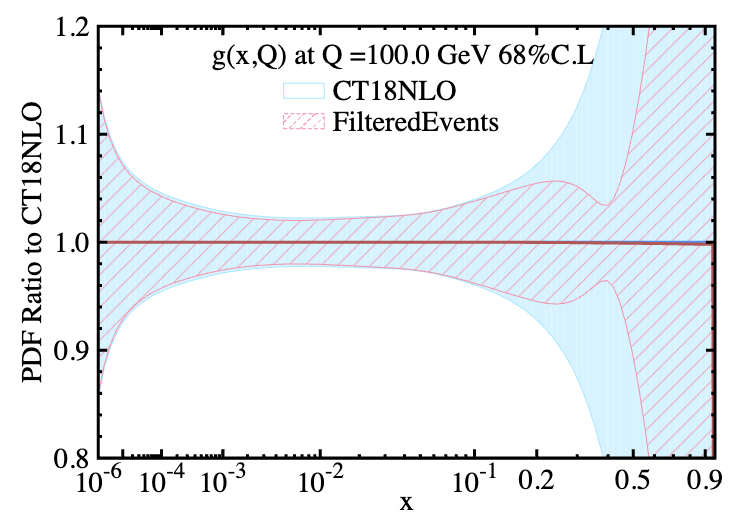}\hfill
    \includegraphics[width=0.3\textwidth]{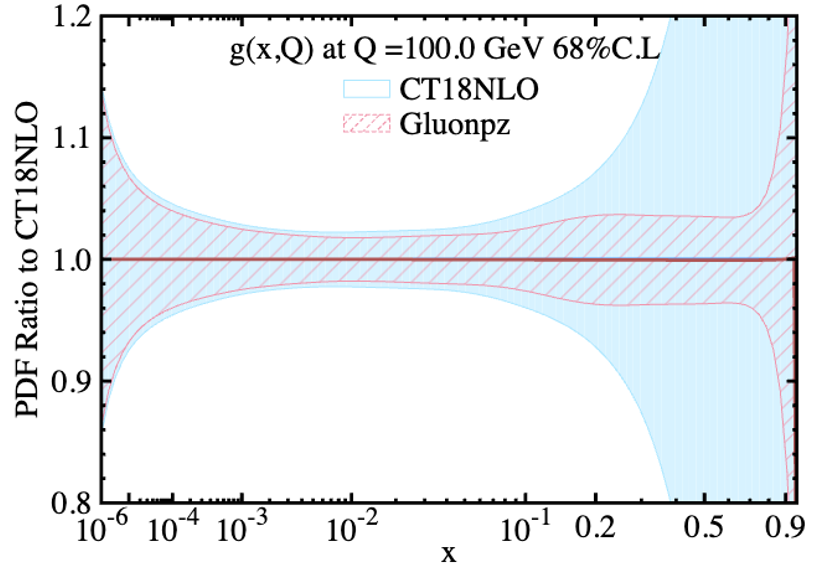}
    \caption{Updated CT18NLO PDF uncertainty bands before and after ePump updates. Left: Updating with the top rapidity distribution from every event. Center: Updating with the top rapidity distribution with events that pass the MLP filter. Right: Updating with the gluon PDF.}
    \label{fig:pdf_updates}
\end{figure}

It can be seen in Fig.~\ref{fig:pdf_updates} that the uncertainty in the high-${x}$ gluon region of the PDF set is heavily reduced when filtering  events.

The updated PDF uncertainty band of the best-case-scenario shows quite dramatic improvements in the gluon PDF uncertainty bands as expected. The uncertainty band can be narrowed up to x=0.9. This shows that there is an opportunity to improve the gluon PDF set with HL-LHC data using machine learning techniques up to a very high parton momentum fraction. 

\section{Outlook and Conclusions}
Other neural network architectures are currently being explored such as graph neural networks, and different techniques like neural network regression. It is currently not well understood how plausible it is to predict the original gluon parton momentum from final state variables or even from reconstruction level variables. This will be explored in future studies. 

This study shows that there is potential to reduce PDF uncertainties by forming variables with machine learning techniques because traditional techniques do not include the full information available for a given process.



\Acknowledgements
This project was supported in part by the National Science Foundation under grant no. PHY-2012165

\bibliography{eprint}{}
\bibliographystyle{unsrt}
 
\end{document}

%% file: econfmacros.tex
\def\beq{\begin{equation}}
\def\eeq#1{\label{#1}\end{equation}}
\def\eeqn{\end{equation}}

\def\beqa{\begin{eqnarray}}
\def\eeqa#1{\label{#1}\end{eqnarray}}
\def\eeqan{\end{eqnarray}}

\let\bar=\overbar

\def\Dslash{\not{\hbox{\kern-4pt $D$}}}
\def\dslash{\not{\hbox{\kern-2pt $\del$}}}

\def\msb{{\bar{\ssstyle M \kern -1pt S}}}

